\documentclass[prl,aps,twocolumn,showpacs,floatfix]{revtex4}

\usepackage{dcolumn}
\usepackage{amssymb,amsmath,amsthm}
\usepackage{graphicx}
\usepackage{epsf,epsfig}
\usepackage{color}

\newcommand \be {\begin{equation}}
\newcommand \ee {\end{equation}}
\newcommand \bea {\begin{eqnarray}}
\newcommand \eea {\end{eqnarray}}
\newcommand \ve {\varepsilon}

\begin{document}

\title{Dependence of the fluctuation-dissipation temperature on the choice of observable}

\author{Kirsten Martens$^1$, Eric Bertin$^2$, and Michel Droz$^1$}
\affiliation{$^1$ Department of Theoretical Physics, University of Geneva,
CH-1211 Geneva 4, Switzerland\\
$^2$ Universit\'e de Lyon, Laboratoire de Physique, ENS Lyon, CNRS,
46 All\'ee d'Italie, F-69007 Lyon, France}
\date{\today}

\begin{abstract}
On general grounds,
a nonequilibrium temperature can be consistently defined from generalized
fluctuation-dissipation relations only if it is independent of the observable
considered. We argue that the dependence on the choice of observable
generically occurs when the phase-space
probability distribution is non-uniform on constant energy shells.
We relate quantitatively this observable dependence to
a fundamental characteristics of nonequilibrium systems,
namely the Shannon entropy difference with respect to the equilibrium
state with the same energy. This relation is illustrated on
a mean-field model in contact with two heat baths at different temperatures.
\end{abstract}


\maketitle

Characterizing nonequilibrium states through generalized, or effective,
thermodynamic parameters is one of the important open issues in
nonequilibrium statistical physics \cite{Jou}.
One possible approach is to introduce thermodynamic parameters conjugated
to conserved quantities \cite{Edwards,Bertin-ITP}.
An alternative approach, more suitable for the definition of temperatures,
is to generalize the fluctuation-dissipation relations (FDR)
that relate the linear response to an external perturbing field
with the correlation of spontaneous fluctuations \cite{Agarwal}.
At equilibrium, the proportionality coefficient
is precisely the temperature. In nonequilibrium situations, one may use
the ratio between correlation and response as a definition
of a non-equilibrium temperature, as proposed in the context of
hydrodynamic turbulence \cite{Hohenberg},
spin-glasses \cite{CuKu93,CuKuPe,Crisanti},
granular materials \cite{Barrat00,Kurchan,Levine},
or sheared fluids \cite{Berthier00}.
The FDR approach is particularly useful since it allows for
experimental \cite{Israeloff,Ciliberto,Ocio,Danna,Joubaud,Gomez}
and numerical \cite{Kob,Makse,Sciortino,Berthier02}
measurements.
A necessary condition for a consistent definition of a FDR-based temperature
is that its value does not depend on the choice of
observables underlying the FDR.
For glassy systems, it has been shown that no dependence on the observable
appears at the mean-field level \cite{CuKu93,CuKuPe}.
This property was also reported in numerical tests
on more realistic models \cite{Berthier02}, although
the conclusions may depend on the model considered \cite{Crisanti}.
For non-glassy systems driven into a nonequilibrium steady-state,
the situation remains unclear, and
no generic conclusion has been reached, even though a lot of work
has been devoted to the study of generalized FDR
\cite{Baldassarri,Bertin-temp,Sasa,Seifert06,Corberi,Cugliandolo07,Maes}.
Theoretical arguments on the observable dependence of the
FDR-temperature are thus highly desirable.

In this Letter, we study how such a dependence on the observable
emerges in a specific class of stationary nonequilibrium systems.
We study the time-dependent linear response 
of a family of observables to an external field.
Relating these response functions to the associated correlation functions
provides us with a set of FDR and with the corresponding effective
temperatures. These temperatures are found to depend on the observable,
a property that we trace back to the non-uniformity of the phase-space
distribution, measured with the Shannon entropy.
A quantitative relation between observable dependence and Shannon entropy
difference with a reference equilibrium state is obtained in a low forcing
limit. We illustrate these results on a fully-connected model
in contact with two heat baths at different temperatures.

Considering a generic system described by a set of $N$ variables $x_i$,
$i=1,\ldots,N$, we introduce a family of observables
$B_p = \sum_{i=1}^N x_i^{2p+1}$, with $p\ge 0$ an integer number.
A small external field $h$, conjugated to an observable $M$, can be applied
to probe the system. The linear response of $B_p$ to the external field
is defined according to the following protocol.
The field $h$ takes a small non-zero value at times $t<0$, and we assume
that the steady state is established. At time $t=0$ the field is switched
off and the time evolution of the observable $B_p$ is recorded. 
The linear response $\chi_p(t)$ is
defined as $\chi_p(t)=\partial \langle\!\langle B_p(t)\rangle\!\rangle/\partial h|_{h=0}$,
the average $\langle\!\langle \cdots \rangle\!\rangle$ being taken over the dynamics resulting from the above protocol.
A FDR holds when the response
$\chi_p(t)$ is proportional to the correlation function (computed for $h=0$)
\be
C_p(t)=\langle (B_p(t)-\langle B_p\rangle)\,(M(0)-\langle
M\rangle)\rangle\;,
\ee
namely
\begin{equation} \label{FDR-p}
\chi_p(t) = \frac{1}{T_p}\, C_p(t)\;.
\end{equation}
The proportionality factor is the inverse of the effective temperature $T_p$, 
which could a priori depend on $p$, and thus on the observable.
Formally, $\langle\!\langle B_p(t)\rangle\!\rangle$ can be expressed as
\be \label{obs}
\langle\!\langle B_p(t)\rangle\!\rangle = \int \prod_{i=1}^N dx_i dx_i'\, B_p \,
G_t^0(\{x_i\}|\{x_i'\}) \mathcal{P}(\{x_i'\},h) 
\ee
with $B_p=B_p(\{x_i\})$, and
where $G_t^0(\{x_i\}|\{x_i'\})$, the zero-field propagator, denotes the
conditional probability to be in a microstate
$\{x_i\} \equiv \{x_i,i=1,\ldots,N\}$ at time $t$
given that the system was in a microstate $\{x_i'\}$ at time $t=0$,
in the absence of the probe field.
The distribution $\mathcal{P}(\{x_i'\},h)$ is the stationary distribution
of the microstate $\{x_i'\}$ in the presence of the field $h$.
Taking the derivative of Eq.~(\ref{obs}) with respect to $h$ at $h=0$,
and using the relation $\partial \mathcal{P}/\partial h = \mathcal{P}\partial \ln \mathcal{P}/\partial h$, we get
\be \label{response}
\chi_p(t) = \left\langle B_p(t)\,
\frac{\partial \ln\mathcal{P}}{\partial h}(\{x_i(0)\},0)\right\rangle\;,
\ee
the average being computed at zero field
\cite{Villamaina,Prost,Seifert09}.

To proceed further, an explicit form of the distribution
$\mathcal{P}(\{x_i\},h)$ is required.
As a general framework, we consider a class of stochastic markovian models,
where a conserved energy $E=\sum_{i=1}^N \ve_h(x_i)$
is randomly exchanged between the internal
degrees of freedom and with the environment (e.~g. reservoirs or external
forces). The external sources and sinks drive the system into a 
nonequilibrium steady state. The resulting
drive can be encompassed by a dimensionless parameter $\gamma$
(e.g.~a normalized temperature difference or external force).
For zero driving, the dynamics satisfies detailed balance and the system
is at equilibrium, characterized by the Gibbs distribution
$\mathcal{P}_\mathrm{eq}(\{x_i\},h)=Z_N^{-1}\exp[-\beta \sum_{i=1}^N \ve_h(x_i)]$ where $\beta=1/T$ is the inverse temperature imposed by an external bath,
and $Z_N$ is the normalization factor.

To simplify the calculations, we assume that the degrees of freedom are
statistically independent, namely
$\mathcal{P}(\{x_i\},h) = \prod_{i=1}^N p(x_i,h)$.
We thus focus on the single-variable probability distribution $p(x,h)$.
Considering the small driving limit $|\gamma| \ll 1$, we expand
the steady-state distribution $p(x,h)$ around
the equilibrium distribution
$p_\mathrm{eq}(x,h) = Z_1^{-1}\exp[-\beta \ve_h(x)]$ as
\begin{equation} \label{dist-dev}
p(x,h) = p_\mathrm{eq}(x,h) \left[1+\gamma F(\ve_h(x))+\mathcal{O}(\gamma^2)\right]\;.
\end{equation}
The normalization of $p(x,h)$ and $p_\mathrm{eq}(x,h)$
imposes $\langle F(\ve)\rangle_\mathrm{eq}=0$,
where $\langle \cdots \rangle_\mathrm{eq}$ is the equilibrium average,
and $\ve$ denotes $\ve_h(x)$.
Note that if $p(x,h)$ follows Eq.~(\ref{dist-dev}), the factorized
$N$-body distribution $\mathcal{P}(\{x_i\},h)$ is in general no longer
a function of the total energy $E=\sum_{i=1}^N \ve_h(x_i)$,
and is thus not uniform over the shells of constant energy.
As a result, the Shannon entropy of the nonequilibrium state should be
lower than the entropy of the equilibrium state with the same energy. 
Hence the entropy difference between the equilibrium and nonequilibrium
states with the same average energy provides an interesting
characterization of the deviation from equilibrium.
Practically, the entropy difference is determined as follows.
We compute the average energy $E(\beta,\gamma)$ of the
out-of-equilibrium system, and we find the temperature $\beta^*$ such
that $E(\beta,\gamma)=E_\mathrm{eq}(\beta^*)$, where
$E_\mathrm{eq}(\beta^*)$ is the equilibrium energy at temperature $\beta^*$.
From the factorization property of $\mathcal{P}(\{x_i\},h)$,
the Shannon entropy
of the whole system is the sum of the entropies associated to each variables
$x_i$, so that we only need to compute the Shannon entropy per degree
of freedom $S=-\int dx\, p(x,h)\ln p(x,h)$.
The entropy difference $\Delta S$ is then defined as
$\Delta S = S_\mathrm{eq}(\beta^*) - S(\beta,\gamma)$,
where $S_\mathrm{eq}(\beta^*)$ is the equilibrium entropy at temperature
$\beta^*$, and $S(\beta,\gamma)$ is the Shannon entropy of the nonequilibrium
state in the presence of a forcing $\gamma$.
After some algebra, one finds

\be\label{ds}
\Delta S = \frac{\gamma^2}{2} \left( \left< F\left(\ve\right)^2 \right>_\mathrm{eq}
- \frac{\langle \ve F\left(\ve\right) \rangle_\mathrm{eq}^2}{\langle \ve^2 \rangle_\mathrm{eq} - \langle \ve \rangle_\mathrm{eq}^2} \right)\;.
\ee
Note that the $\gamma^2$-term in the expansion (\ref{dist-dev})
of $p(x,h)$ needs to be taken into account in the calculation,
but eventually cancels out.
It can be checked that $\Delta S\ge 0$, although this property is not explicit
in Eq.~(\ref{ds}).
The equality $\Delta S = 0$ is obtained for a linear $F(\ve)$,
as in this case, $p(x,h)$ can be recast into an equilibrium form with an
effective temperature --see Eq.~(\ref{dist-dev}).
Turning to the non-linear case, we parameterize $F(\ve)$ as
$F(\ve) = a+b\ve+\eta f(\ve)$, where $\eta$ characterizes the
amplitude of the nonlinearity. The parameter $a$ is fixed by the constraint
$\langle F(\ve)\rangle=0$. We then find $\Delta S=\gamma^2 \eta^2 \sigma$,
where $\sigma$ is a constant depending on the detailed shape
of the functions $f(\ve)$ and $\ve_h(x)$.
For instance, in the case $f(\ve)=\ve^2$ with $h=0$ and
$\ve_0(x)=\frac{1}{2}x^2$, one has
$\Delta S = \frac{3}{4} \gamma^2 \eta^2/\beta^4$.

We now come back to the FDR. As a simplifying hypothesis,
we assume that the dynamics is such that each event decorrelates the involved
variables $x_i$ from their previous values (see below for an explicit example).
Hence all correlation functions are proportional to
the persistence probability $\Phi(t)$, that is the probability
that no event involving a given variable $x_i$ occurred between times $0$
and $t$ \cite{Bertin-temp}.
Expanding for small field $h$ the local energy
$\ve_h(x)=\ve_0(x)-h\psi(x) + \mathcal{O}(h^2)$,
with $\psi(x)$ an odd function,
we find $C_p(t)=N\langle x(t)^{2p+1} \psi(x(0)) \rangle
= N \langle x^{2p+1} \psi(x)\rangle\, \Phi(t)$, the last average
being performed on the steady-state distribution.
A similar calculation starting from Eq.~(\ref{response}) yields
for the response function
\be
\chi_p(t)=N[\beta \langle x^{2p+1}\psi(x)\rangle -
\gamma \langle x^{2p+1}\psi(x) F'(\ve_0) \rangle]\, \Phi(t)
\ee
where $\ve_0$ stands for $\ve_0(x)$.
To first order in $\gamma$, the average in the second term
can be replaced by the equilibrium average.
Expressing $\Phi(t)$ as a function of $C_p(t)$, we find that
the FDR (\ref{FDR-p}) is obeyed, with $\beta_p = T_p^{-1}$ given by

\be\label{beta_p}
\beta_p = \beta-\gamma \frac{\langle x^{2p+1}\psi(x) F'(\ve_0)\rangle_\mathrm{eq}}{\langle x^{2p+1} \psi(x)\rangle_\mathrm{eq}}\;.
\ee
As expected, $T_p$ generically depends on $p$, that is on the
observable. Yet, a linear $F(\ve)$, namely $F(\ve)=a+b\ve$,
yields an effective temperature $\beta_p=\beta-\gamma b$ that does not
depend on the observable.
A dependence on the observable arises when $F(\ve)$ has a nonlinear
contribution. Using the parameterization
$F(\ve) = a+b\ve+\eta f(\ve)$, we get from Eq.~(\ref{beta_p})
that $\beta_p-\beta_{p=0}$ is proportional to $\eta$.
Hence the dependence of the fluctuation-dissipation temperature on the
observable is directly related to the amplitude of the non-linearity
in $F(\ve)$. As this amplitude $\eta$ is also captured by the entropy
difference $\Delta S$, it is interesting to relate quantitatively
$\beta_p-\beta_0$ to $\Delta S$. We obtain

\be \label{observable-entropy}
\frac{|\beta_p-\beta_0|}{\beta} = \kappa_p \sqrt{\Delta S}\;,
\ee
where $\kappa_p>0$ is a dimensionless constant,
depending on $p$, on the functional forms of $f(\ve)$
and $\ve_h(x)$, but not on $\gamma$ and $\eta$.
As an example, one finds in the case $f(\ve)=\ve^2$ and
$\ve_h(x)=\frac{1}{2}x^2-hx+\mathcal{O}(h^2)$ that $\kappa_p=4p/\sqrt{3}$.
Thus it turns out that the dependence of the fluctuation-dissipation
temperature on the choice of observable is a direct measure of the
deviation from equilibrium.

\begin{figure}[b]
\centering\includegraphics[width=0.7\columnwidth,clip]{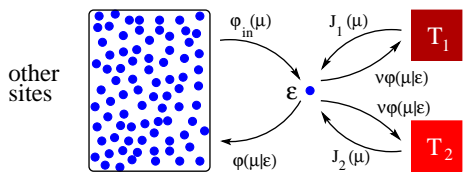}
\caption{(Color online) Scheme of the mean-field model. A single site contains an amount of energy $\varepsilon$. It is in contact with two baths at different
temperatures $T_1=\beta_1^{-1}$ and $T_2=\beta_2^{-1}$
and with the other sites.}
\label{fig-mf-model}
\end{figure}

To illustrate this result, we consider an
energy transport model on a fully connected geometry,
in contact with two heat baths at
inverse temperatures $\beta_1$ and $\beta_2$.
The contact is characterized by a coupling constant $\nu$.
A random fraction $\mu$ of the local energy
$\ve_i \equiv \ve_h(x_i) = \frac{1}{2} (x_i-h)^2$
is transferred from a site $i$ to an arbitrary site $j$ with
a probability rate $\varphi(\mu|\ve_i)$.
After the transfer, $x_i$ and $x_j$ are changed according to
$x_i'-h\!=\!\pm\sqrt{(x_i-h)^2-2\mu}$ and $x_j'-h\!=\!\pm\sqrt{(x_j-h)^2+2\mu}$
with equiprobable and uncorrelated random signs.
The heat baths are characterized by injection rates
$J_{\alpha}(\mu)= \nu e^{-\beta_{\alpha}\mu}$ ($\alpha=1,2$),
while transfers from the site $i$ to any of the heat baths follow the rate
$\nu \varphi(\mu|\ve_i)$.
We choose $\varphi(\mu|\ve_i)=(1-\mu/\ve_i)^{-1/2}$ to ensure that
the equilibrium distribution is recovered when $\beta_1=\beta_2$
\cite{Evans}.
A sketch of the dynamics is shown on Fig.~\ref{fig-mf-model}.
Due to the fully connected geometry of the model,
the different sites become statistically independent
in the thermodynamic limit $N \to \infty$, so that the single-site
distribution $p(x,h)$ provides an exact description in this limit.
In addition, as the dynamics involves only redistributions of the local energy
$\ve_i=\frac{1}{2}(x_i-h)^2$, the model can be effectively described
in terms of local energies. The stationary distribution $p(x,h)$
is then completely determined by the distribution $P(\ve)$.
This change of variables is expressed in the probability 
densities as
$p(x,h)=\frac{1}{2}|x-h|P(\frac{1}{2}(x-h)^2)$.
The equilibrium distribution at temperature $T=\beta^{-1}$,
$p_\mathrm{eq}(x,h)=\sqrt{\beta /2\pi}\exp[-\beta (x-h)^2/2]$
corresponds to
$P_\mathrm{eq}(\ve)=\sqrt{\beta/\pi\ve}\;\exp(-\beta\ve)$. 
The master equation for the time-dependent $N$-site probability distribution
can then be recast into a nonlinear evolution equation for
the single-site energy distribution $P(\varepsilon,t)$:
\begin{eqnarray}\label{P-evolution}
\frac{\partial P}{\partial t}(\varepsilon,t) &=& \int_0^\varepsilon d\mu\, [J_1(\mu)+J_2(\mu)+\varphi_\mathrm{in}(\mu,t)] P(\varepsilon-\mu,t)\nonumber\\
&& -\int_0^\infty d\mu\, [J_1(\mu)+J_2(\mu)+\varphi_\mathrm{in}(\mu,t)] P(\varepsilon,t)\nonumber\\
&& +(2\nu+1) \int_0^\infty d\mu\, \varphi(\mu|\varepsilon+\mu)P(\varepsilon+\mu,t)\nonumber\\
&& -(2\nu+1) \int_0^\varepsilon d\mu\, \varphi(\mu|\varepsilon)P(\varepsilon,t)
\end{eqnarray}
where $\varphi_\mathrm{in}(\mu,t)=\int_\mu^\infty d\varepsilon\, \varphi(\mu|\varepsilon)P(\varepsilon,t)$
accounts for the energy transfers coming from all the other sites.

\begin{figure}[t]
\centering\includegraphics[width=0.8\columnwidth,clip]{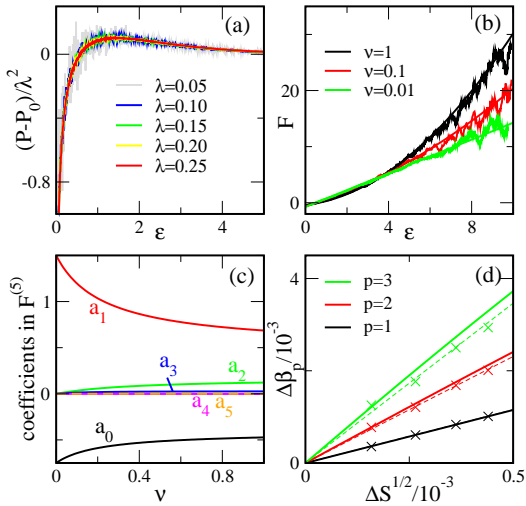}
\caption{(Color online) (a) $(P-P_0)/\lambda^2$ as a function of the local energy $\ve$ obtained by numerical simulations of the dynamics ($\nu=1$, $T_\mathrm{max}=10^8$). (b) The function $F(\ve)$ obtained by simulations (noisy lines) in comparison with the analytically obtained results for $F^{(2)}(\ve)$ (solid lines) for different values of $\nu$ ($\lambda=0.2$, $T_\mathrm{max}=10^7$). (c) $\nu$-dependence of the coefficients in $F^{(5)}(\ve)$ (see text). (d) Parametric plot in $\nu$ of $\Delta \beta_p=|\beta_p-\beta_0|/\beta$ versus $\sqrt{\Delta S}$ obtained using $F^{(5)}(\ve)$, either fitted to numerical data ($\times$) or calculated in the analytical approximation (solid lines).
Dashed lines: Eq.~(\ref{observable-entropy}) with $\kappa_p=4p/\sqrt{3}$.
Simulation parameters: $\lambda=0.05$, $T_\mathrm{max}=10^7$.}
\label{fig-F2}
\end{figure}

To determine the steady-state distribution $P(\ve)$,
we consider the limit of a small temperature difference
$\beta_1=\beta (1-\lambda)$ and $\beta_2=\beta (1+\lambda)$,
with $\lambda \ll 1$, and expand the distribution in $\lambda$.
The linear term in $\lambda$ vanishes because the two heat baths 
play a symmetric role. The leading correction should thus behave as
$\lambda^2$, so that the distribution can be written in a form
similar to Eq.~(\ref{dist-dev}), namely
\be
P(\ve) = P_\mathrm{eq}(\ve) \left[1+\lambda^2 F(\ve)
+\mathcal{O}(\lambda^4)\right]\;.
\ee
This scaling form is validated by direct numerical simulations
of the stochastic dynamics, as shown in Fig.~\ref{fig-F2}(a).
Simulations were performed on a system of size $N=102$, with $\beta=1$.
The relatively small system size allows for long time averaging until a time
$T_\mathrm{max} \sim 10^7$ or $10^8$, so as to reach a satisfactory statistics
(in a unit of time, all sites have in average experienced about one
redistribution event). In Fig.~\ref{fig-F2}(b),
the function $F(\varepsilon)$ obtained from simulations 
is shown for different values of the coupling
strength $\nu$. We observe that the curvature of $F(\ve)$ is reduced
when decreasing $\nu$. Eq.~(\ref{P-evolution}) has no exact solution
involving a finite polynomial function $F(\ve)$.
To find the best polynomial approximation
$F^{(L)}(\ve) = \sum_{k=0}^L a_k \beta^k \ve^k$
at a given order $L$, we devised a variational procedure
($\beta$ factors are introduced to make $a_k$ dimensionless).
The function $F^{(L)}(\ve)$ is obtained analytically by minimizing the error,
under the constraints of normalization and zero neat flux with the baths, in the evolution equation (\ref{P-evolution}) linearized in $F(\ve)$.
The error is defined as the equilibrium average of the square
of the r.h.s.~in the linearized equation.
For $L>2$, we find that the coefficients $a_k$, $k>2$, in the expansion
are numerically small, as illustrated in Fig.~\ref{fig-F2}(c).
A second order polynomial is thus already a good approximation of $F(\ve)$ for $\nu\lesssim 1$ [see Fig.~\ref{fig-F2}(b)].
Taking into account higher order terms in $F(\ve)$, we find that
the relation (\ref{observable-entropy}) between the observable
dependence and the entropy difference is valid to a good accuracy
[Fig.~\ref{fig-F2}(d)].
In addition, we observe that in the limit $\nu \to 0$,
the coefficients $a_k$, $k>1$ vanish while $a_0 \to -\frac{3}{4}$
and $a_1 \to \frac{3}{2}$. Therefore the temperature $T_p$
becomes observable independent in the small coupling limit.

To sum up, we have shown that a large class of nonequilibrium systems
generically exhibit observable dependence of the fluctuation-dissipation
ratio, even in the mean-field case.
Accordingly, a unique nonequilibrium temperature cannot be defined
from the FDR.
The dependence on the observable can be traced back to the non-uniformity
of the phase-space distribution on shells of constant energy,
quantified by the difference of Shannon entropy between
the equilibrium and nonequilibrium states with the same energy.
We have illustrated these results explicitly on a mean-field model
connected to two heat baths, confirming that observable dependence
appears in the driven stationary state.
The dependence however becomes weaker when the coupling to the reservoirs
is decreased. This might be the reason why the observable dependence
has not been encountered in numerical simulations of granular gases
\cite{Baldassarri}.
Furthermore, it turns out that the entropy difference $\Delta S$
is a relevant characterization of nonequilibrium systems.
If $\Delta S=0$, a single temperature emerges from the FDR,
and the statistical properties resemble closely that of equilibrium systems.
In contrast, if $\Delta S>0$, the system can be described by two parameters,
a reference temperature (e.g., $T_{p=0}$) and $\Delta S$.
It would be interesting to try to measure $\Delta S$ experimentally or
numerically through the use of the FDR, in real
out-of-equilibrium systems like granular
gases \cite{Baldassarri} or turbulent flows \cite{Naert}.



\begin{thebibliography}{99}


\bibitem{Jou}
J. Casas-V\'azquez and D. Jou, Rep. Prog. Phys. {\bf 66}, 1937 (2003).

\bibitem{Edwards}
S.~F. Edwards, {\it Granular Matter: An Interdisciplinary Approach}
(Springer Verlag, New York, 1994).

\bibitem{Bertin-ITP}
E. Bertin, O. Dauchot, and M. Droz, Phys. Rev. Lett. {\bf 96}, 120601 (2006);
E. Bertin, K. Martens, O. Dauchot, and M. Droz, Phys. Rev. E {\bf 75}, 031120
(2007).

\bibitem{Agarwal}
G.~S. Agarwal, Z. Phys. {\bf 252}, 25 (1972).

\bibitem{Hohenberg}
P. Hohenberg, B. Shraiman, Physica D, {\bf 37}, 109 (1989).

\bibitem{CuKu93}
L.~F. Cugliandolo, J. Kurchan, Phys. Rev. Lett. {\bf 71}, 173 (1993).

\bibitem{CuKuPe}
L.~F. Cugliandolo, J. Kurchan, and L. Peliti, Phys. Rev. E {\bf 55}, 3898
(1997).

\bibitem{Crisanti}
A. Crisanti and F. Ritort, J. Phys. A. {\bf 36}, R181 (2003).

\bibitem{Barrat00}
A. Barrat, J. Kurchan, V. Loreto, and M. Sellitto, Phys. Rev. Lett. {\bf 85}, 5034 (2000).

\bibitem{Kurchan}
J. Kurchan, J. Phys.: Cond. Matt. {\bf 12}, 6611 (2000);
Nature {\bf 433}, 222 (2005).

\bibitem{Levine}
Y. Shokef, G. Bunin, and D. Levine, Phys. Rev. E {\bf 73}, 046132 (2006);
G. Bunin, Y. Shokef, and D. Levine, Phys. Rev. E {\bf 77}, 051301 (2008).

\bibitem{Berthier00}
J.-L. Barrat and L. Berthier, Phys. Rev. E {\bf 63}, 012503 (2000).

\bibitem{Israeloff}
T.~S. Grigera and N.~E. Israeloff, Phys. Rev. Lett. {\bf 83}, 5038 (1999).

\bibitem{Ciliberto}
L. Bellon, S. Ciliberto, and C. Laroche, Europhys. Lett. {\bf 53}, 511 (2001).

\bibitem{Ocio}
D. H\'erisson and M. Ocio, Phys. Rev. Lett. {\bf 88}, 257202 (2002).

\bibitem{Danna}
G. D'Anna {\it et. al.}, Nature {\bf 424}, 909 (2003).

\bibitem{Joubaud}
S. Joubaud {\it et.~al.}, Phys. Rev. Lett. {\bf 102}, 130601 (2009).

\bibitem{Gomez}
J.~R. Gomez-Solano {\it et.~al.}, Phys. Rev. Lett. {\bf 103}, 040601 (2009).

\bibitem{Kob}
J.-L. Barrat and W. Kob, Europhys. Lett. {\bf 46}, 637 (1999).

\bibitem{Makse}
H.~A. Makse and J. Kurchan, Nature {\bf 415}, 614 (2002).

\bibitem{Sciortino}
F. Sciortino and P. Tartaglia, Phys. Rev. Lett. {\bf 86}, 107 (2001).

\bibitem{Berthier02}
L. Berthier and J.-L. Barrat, Phys. Rev. Lett. {\bf 89}, 095702 (2002);
J. Chem. Phys. {\bf 116}, 6228 (2002).

\bibitem{Baldassarri}
A. Puglisi, A. Baldassarri, and V. Loreto, Phys. Rev. E 
{\bf 66}, 061305 (2002).

\bibitem{Bertin-temp}
E. Bertin, O. Dauchot and M. Droz, Phys. Rev. Lett. {\bf 93}, 230601 (2004);
Phys. Rev. E {\bf 71}, 046140 (2005).

\bibitem{Sasa}
T. Harada and S.-i. Sasa, Phys. Rev. Lett. {\bf 95}, 130602 (2005).

\bibitem{Seifert06}
T. Speck and U. Seifert, Europhys. Lett. {\bf 74}, 391 (2006).

\bibitem{Corberi}
F. Corberi, E. Lippiello, and M. Zannetti, J. Stat. Mech. P07002 (2007).

\bibitem{Cugliandolo07}
D. Loi, S. Mossa, and L.~F. Cugliandolo, Phys. Rev. E {\bf 77}, 051111 (2008).

\bibitem{Maes}
M. Baiesi, C. Maes, and B. Wynants, Phys. Rev. Lett. {\bf 103}, 010602 (2009).

\bibitem{Villamaina}
D. Villamaina, A. Baldassarri, A. Puglisi and A. Vulpiani,
J. Stat. Mech. P07024 (2009).

\bibitem{Prost}
J. Prost, J.-F. Joanny and J.~M.~R. Parrondo, Phys. Rev. Lett.
{\bf 103}, 090601 (2009).

\bibitem{Seifert09}
U. Seifert and T. Speck, arXiv:0907.5478.


\bibitem{Evans}
M.~R. Evans, S.~N. Majumdar, and R.~K.~P. Zia, J. Phys. A \textbf{37},
L275 (2004).

\bibitem{Naert}
V. Grenard, N.~B. Garnier and A. Naert, J. Stat. Mech. L09003 (2008).


\end{thebibliography}
\end{document}